\documentstyle[12pt]{article}
\newcommand{\beq}{\begin{equation}}
\newcommand{\eeq}{\end{equation}}
\newcommand{\id}{i\!\!\not\!\partial}
\newcommand{\as}{\not\! a}
\newcommand{\bs}{\not\! b}

\newcommand{\D}{{\cal{D}}}
\newcommand{\dv}{d^3x}
\newcommand{\Z}{{\cal Z}}

\begin{document}
\title
{The Fermion-Boson Mapping in Three Dimensional Quantum Field Theory}
\author{Eduardo Fradkin\\
{\normalsize \it Department of Physics,
University of Illinois at Urbana-Champaign}\\
{\normalsize \it 1110 W.Green St., Urbana, Illinois 61801-3080, USA}\\
\rule{0cm}{1.cm}
{\normalsize \it and}\\
\rule{0cm}{1.cm}
Fidel A. Schaposnik\thanks{Investigador CICBA, Argentina}\\
{\normalsize\it Departamento de F\'\i sica, Universidad
Nacional de La Plata}\\ {\normalsize\it C.C. 67, (1900) La Plata,
Argentina}\\
}
\date{}

\maketitle

\def\thepage{\protect\raisebox{0ex}{\ } La Plata 94-08}
\thispagestyle{headings}
\markright{\thepage}

\begin{abstract}
{\normalsize {We discuss bosonization in three dimensions by
establishing a connection between the massive Thirring model
and the Maxwell-Chern-Simons theory. We show, to lowest order
in inverse fermion mass, the identity between the corresponding
partition functions; from this, a bosonization identity for the
fermion current, valid for length scales long compared with the
Compton wavelength of the fermion, is inferred. We present a
non-local operator in the Thirring model which exhibits
fractional statistics. }}
\end{abstract}
\newpage
\pagenumbering{arabic}

In this paper we investigate the problem of the mapping of
quantum field theories of interacting fermions in $2+1$
dimensions onto an equivalent theory of interacting bosons.
These mappings, commonly called {\it bosonization}, are well
established in the context of $1+1$ dimensional theories.
There, bosonization  constitutes one of the main tools
available for the study of the non-perturbative behavior of
both quantum field theories \cite{coleman} and of condensed
matter systems \cite{mattis}. The bosonization identities,
which relate the fermionic current with the topological current
of a bosonic theory, can be viewed as a consequence of a
non-trivial current algebra. However, in dimensions other than
$1+1$, much less is known. Although current algebras do exist
in all dimesions, the Schwinger terms have a much more complex
structure in higher dimensions. Also, a simple counting of
degrees of freedom shows that a simple minded mapping between
fermions and {\it scalars} can only hold in $1+1$ dimensions.
In this work we show that the {\it low energy sector} of a
theory of massive self-interacting fermions, the massive
Thirring Model in $2+1$ dimeensions, can be bosonized. However,
the bosonized theory is a {\it gauge theory}, the
Maxwell-Chern-Simons gauge theory.

Some time ago Deser and Redlich \cite{DR} discussed the
equivalence of the three dimensional effective electromagnetic
action of the $CP^1$ model and of a charged  massive fermion to
lowest order in inverse (fermion) mass, following the ideas of
Refs. \cite{Pol}-\cite{Wig}. This issue  is relevant in the
context of transmutation of spin and statistics in three
dimensions, with applications to interesting problems both in
Quantum Field Theory and Condensed Matter physics \cite{Ed}.
The mapping, first discussed by Polyakov\cite{Pol} and extended
by Deser and Redlich\cite{DR}, shows that a massive scalar
particle coupled to a Chern-Simons gauge field becomes a
massive Dirac fermion for a properly chosen value of the
Chern-Simons coupling. As such, this Bose-Fermi transmutation
is a property which holds only at very long distances, {\it
i.e.} at scales long compared with the Compton wavelength of
the particle. In terms of the path-integral picture, the
transmutation is a property of very large, smooth, paths.
Hence,  these results hold to lowest order in an expansion in
powers of the mass of the particle. Other approaches to
bosonization in $2+1$ dimensions have been developed in Refs.
\cite{marino}-\cite{kovner}.

In the same vein as in \cite{DR}, we establish here, to leading
order in the inverse fermionic mass, an identity between the
partition functions for the three-dimensional Thirring model
and the topologically massive $U(1)$ gauge theory, the
Maxwell-Chern-Simons theory (MCS). This result enlarges the
boson-fermion correspondence by connecting a self-interacting
fermion model  and a Chern-Simons system.

We start from the three-dimensional (Euclidean) massive
Thirring model Lagrangian:

\beq
{\cal L}_{Th}= \bar\psi^i (\id + m) \psi^i -\frac{g^2}{2N}j^{\mu}j_{\mu}
\label{1}
\eeq
where $\psi^i$ are N two-component Dirac spinors and $J^{\mu}$ the $U(1)$
current,
\beq
j^{\mu}=\bar\psi^i\gamma^{\mu}\psi^i.
\label{2}
\eeq
The coupling constant $g^2$ has dimensions of inverse mass.
(Although non-renormalizable by power counting, four fermion
interaction models in $2+1$ dimensions are known to be
renormalizable in the $1/N$ expansion \cite{Gross}.)

The partition function for the theory is defined as
\beq
\Z_{Th} = \int \D\bar\psi \D\psi \exp[
-\int \left ( \bar\psi^i (\id + m) \psi^i -\frac{g^2}{2N}J^{\mu}J_{\mu}
\right ) \dv]
\label{3}
\eeq

We now  eliminate the quartic
interaction by introducing a vector field $a_{\mu}$ through the identity
\beq
\exp(\int \frac{g^2}{2N}J^{\mu}J_{\mu} \dv)=\int\D a_{\mu}
\exp[-\int (\frac{1}{2}a^{\mu}a_{\mu}+\frac{g}{\sqrt{N}}J^{\mu}a_{\mu})\dv]
\label{4}
\eeq
(up to a multiplicative normalization constant) so that the partition
function becomes
\beq
\Z_{Th} =  \int \D\bar\psi \D\psi \D a_{\mu} \exp[
-\int(\bar\psi^i (\id + m + \frac{g}{\sqrt{N}}\as)\psi^i
+\frac{1}{2} a^{\mu}a_{\mu})\dv].
\label{5}
\eeq

We are now going to perform the fermionic path-integral which
gives as usual the Dirac operator determinant:

\beq
\int \D\bar\psi \D\psi \exp(-\int\bar\psi^i (\id + m +
\frac{g}{\sqrt{N}}\as)\psi^i \dv) = \det (\id+ m +\frac{g}{\sqrt{N}}\as)
\label{8}
\eeq
Being the Dirac operator unbounded, its determinant requires
regularization. Any sensible regularization approach (for
example, $\zeta$-function or Pauli-Villars approaches) gives a
parity violating contribution \cite{NS}-\cite{GMSS}. There are
also parity conserving terms which have been computed as an
expansion in inverse powers of the fermion mass:

\beq
 \ln \det (\id+ m +\frac{g}{\sqrt{N}}\as) = \pm \frac{ig^2}{16\pi }
\int\epsilon_{\mu\nu\alpha} f^{\mu\nu} a^{\alpha} \dv + I_{PC}[a_{\mu}]+
O(\partial^2/m^2)
,
\label{9f}
\eeq

\beq
 I_{PC}[a_{\mu}] = - \frac{g^2}{24\pi  m} \int \dv  f^{\mu\nu} f_{\mu\nu}
+ \ldots
\label{8f}
\eeq
Using this result we can write $Z_{Th}$ in the form:

\beq
Z_{Th} = \int Da_{\mu}  \exp(-S_{eff}[a_{\mu}])
\label{12f}
\eeq
where $S_{eff}[a_{\mu}]$ is given by
\begin{eqnarray}
S_{eff}[a_{\mu}] &=& \frac{1}{2}\int \dv (a_{\mu}a^{\mu} \mp
\frac{ig^2}{4\pi}\epsilon^{\mu\alpha\nu}a_{\mu}\partial_{\alpha}a_{\nu})  +
\nonumber \\
& +& \frac{g^2}{24\pi m} \int \dv  f^{\mu\nu} f_{\mu\nu} +
O(\partial^2/m^2)
\label{thicon}
\end{eqnarray}


Up to corrections of order $1/m$, we recognize in $S_{eff}$ the
self-dual action $S_{SD}$ introduced some time ago by Townsend,
Pilch and van Nieuwenhuizen \cite{TPv},

\beq
S_{SD} = \frac{1}{2}\int \dv (a_{\mu}a^{\mu} \mp
\frac{ig^2}{4\pi}\epsilon^{\mu\alpha\nu}a_{\mu}\partial_{\alpha}a_{\nu})
\label{SD}
\eeq
Then, to leading order in $1/m$ we have established the
following identification:

\beq
Z_{Th} \approx \int Da_{\mu} \exp(-S_{SD})
\label{nue}
\eeq
Now, Deser and Jackiw \cite{DJ} have proven the equivalence
between the model with dynamics defined by $S_{SD}$ and the
Maxwell-Chern-Simons theory . In what follows, we shall adapt
Deser-Jackiw arguments to the path-integral framework showing
that to the leading order in $1/m$ expansion the Thirring model
partition function coincides with that of the MCS theory. To
this end, let us introduce an ``interpolating action" $S_{
I}[a_{\mu}, A_{\mu}]$

\beq
S_{I}[a_{\mu}, A_{\mu}] = \int \dv (\frac{1}{2}a_{\mu}a^{\mu} -
i\epsilon^{\mu\alpha\nu}a_{\mu}\partial_{\alpha}A_{\nu} \mp i
\frac{2\pi}{g^2} \epsilon^{\mu\alpha\nu}A_{\mu}\partial_{\alpha}
A_{\nu} )
\label{Master}
\eeq
and the corresponding partition function  $Z_I$, a path-integral
over both $A_{\mu}$ and $a_{\mu}$,

\beq
Z_{I} = \int DA_{\mu}Da_{\mu} \exp(-S_I)
\label{i}
\eeq
The theory with action $S_I$ is invariant under the local gauge
transformation

\beq
\delta A_{\mu} = \partial_{\mu} \omega, \quad \delta a_{\mu} = 0
\label{t1}
\eeq

To see the connection between $Z_I$ and $Z_{Th}$ when written as in
(\ref{thicon}) and (\ref{SD}), let us perform the path-integral
over $A_{\mu}$ in (\ref{Master}),
\begin{eqnarray}
I[a_{\mu}] &\equiv&\int DA_{\mu} \exp[\int \dv (\pm i\frac{2\pi}{g^2}
\epsilon^{\mu\alpha\nu}A_{\mu}\partial_{\alpha}A_{\nu}-i \epsilon^{\mu
\alpha \nu} a_\mu \partial_\alpha A_\nu)] \nonumber\\
& = &\int DA_{\mu} \exp[-\int \dv \frac{1}{2}A_{\mu}S^{\mu \nu}A_{\nu} +
A_{\mu}J^{\mu}]
\label{soba}
\end{eqnarray}
where we have scaled $A_{\mu} \to \sqrt{4\pi/g^2} A_{\mu}$ and defined

\beq
J^{\mu} = i\sqrt{\frac{g^2}{4\pi}}\epsilon^{\mu \rho \sigma}
\partial_{\rho}a_{\sigma}
\label{jota}
\eeq

\beq
S^{\mu \nu} = \mp i\epsilon^{\mu \alpha \nu} \partial_{\alpha}
\label{S}
\eeq
Being $ S^{\mu \nu}$ non-invertible
we shall define a regulated operator $S^{\mu \nu}[\Lambda]$,

\beq
S^{\mu \nu}[\Lambda] = \mp i\epsilon^{\mu \alpha \nu} \partial_{\alpha}
+\frac{1}{\Lambda}\partial^{\mu}\partial{\nu}
\label{reg}
\eeq
so that $I[a_{\mu}]$ can be calculated from the identity

\beq
I[a_{\mu}] = \lim_{\Lambda \to \infty} I^{\Lambda}[a_{\mu}]
\label{reg11}
\eeq

\beq
I^{\Lambda}[a_{\mu}] = \int DA_{\mu}
\exp \left (-\int \dv (\frac{1}{2}A_{\mu}S^{\mu \nu}[\Lambda]A_{\nu} +
A_{\mu}J^{\mu})\right )
\label{reg2}
\eeq
Now,  $I^{\Lambda}[a_{\mu}]$ can be easily calculated,

\beq
 I^{\Lambda}[a_{\mu}] = \exp(-\int d^3x d^3y J^{\mu}(x)
S^{-1}_{\mu \nu}[\Lambda]  J^{\nu}(y)
\label{lain}
\eeq
with

\beq
S^{-1}_{\mu \nu}[\Lambda] = \pm \frac{i}{4\pi} \epsilon_{\mu \nu
\alpha}\partial^{\alpha} \frac{1}{\vert x-y\vert} - \frac{\Lambda}{8\pi}
\partial_{\mu}\partial_{\nu}\vert x-y\vert
\label{green}
\eeq
so that we finally have

\beq
I[a_{\mu}] = \exp(\pm i \int \dv \frac{g^2}{8\pi}\epsilon^{\mu\alpha\nu}
a_{\mu}\partial_{\alpha}a_{\nu})
\label{SD1}
\eeq
and with this we see that

\beq
 Z_{I} = \int Da_{\mu} \exp(-S_{SD})
\label{p1}
\eeq
Let us point that the non invertibility of the operator $S_{\mu \nu}$ is
a consequence of the gauge invariance of the action $S_I$.
The extra regulating term in (\ref{reg}) can be interpreted as originating
from a gauge fixing term in the action of the form $({\Lambda}/{2})
(\partial_\mu A^\mu)^2$ with the limit $\Lambda \to \infty$ enforcing
the Lorentz gauge $ \partial_\mu A^\mu=0$.

Using eq.(\ref{nue}) we can then establish the following relation:
\beq
Z_{Th} \approx Z_I
\label{ide1}
\eeq
Now, if instead of integrating out $A_{\mu}$ in $Z_I$ we integrate over
$a_{\mu}$, we easily find

\beq
Z_L = \int DA_{\mu} \exp \int \dv (\frac{1}{4} F_{\mu \nu}^2  \pm i
\frac{2\pi}{g^2} \epsilon^{\mu\alpha\nu}A_{\mu}\partial_{\alpha}A_{\nu} )
\label{casi}
\eeq
which is nothing but the partition function $Z_{MCS}$ for the
Maxwell-Chern-Simons theory \cite{Jac2}-\cite{Jac3}. Then, using
eq.(\ref{ide1}) one finally proves the equivalence, to leading
order in $1/m$,
of the partition functions for  Thirring model and  the MCS theory:

\beq
Z_{Th} \approx Z_{MCS}
\label{fin1}
\eeq
(As stressed above, $\approx$ means that the identification has been
proven to leading order in $1/m$)

Equation (\ref{fin1}) is one of the main results in our work.
It expresses the equivalence
between the low energy sector of a theory of $3$-dimensional
fer-mions (interacting via a current-current term) and  (gauge) bosons
(with Maxwell-Chern-Simons action). The Thirring coupling constant
$g^2/N$ in the fermionic model enters as $2\pi/g^2$ in the CS term.
This means that the Thirring spin $1/2$ fermion in $2+1$ dimensions is
equivalent to a spin $1$ massive excitation, with mass
$\pi/g^2$ \cite{Jac2}-\cite{Jac3}.

The equivalence has been established at lowest order in $1/m$.
However, if we follow Deser and Redlich \cite{DR} and consider
that $g^2 = c/m$, with $c$ a dimensionless coupling constant,
the first term in $I_{PC}$ becomes of order $1/m^2$ (see
eq.(\ref{8f})) and the equivalence is then extended to the next
order.

In order to infer the bosonization recipe deriving from the
equivalence, we  add a source for the Thirring current:

\beq
L_{Th}[b_{\mu}] = L_{Th} + \int \dv j^{\mu}b_{\mu}
\label{s1}
\eeq
Then, instead of (\ref{5}), the partition function now reads:

\beq
\Z_{Th}[b_{\mu}] =  \int \D\bar\psi \D\psi \D a_{\mu} \exp[
-\int(\bar\psi^i (\id + m + \frac{g}{\sqrt{N}}\as + \bs)\psi^i
+\frac{1}{2} a^{\mu}a_{\mu})\dv].
\label{s5}
\eeq
Or, after shifting $a_{\mu} \to a_{\mu} - (g/\sqrt{N})b_{\mu}$,

\beq
Z_{Th}[b_{\mu}] = \exp \left (-\frac{N}{2g^2} \int \dv b_{\mu}
b^{\mu}\right) \times \int Da_{\mu}
\exp( -S_{eff}[a_{\mu}] +\frac{\sqrt N}{g^2}\int \dv b_{\mu} a^{\mu})
\label{s6}
\eeq
with $S_{eff}[a_{\mu}]$ still given by eq.(\ref{thicon}). Then,
we can again establish to
order $1/m$ the connection between the Thirring and self-dual models,
in the presence of sources:

\beq
Z_{Th}[b_{\mu}] = \exp \left (-\frac{N}{2g^2} \int
\dv b_{\mu}b^{\mu}\right)\times \int Da_{\mu} \exp(-S_{SD} +
\frac{\sqrt N}{g}\int \dv b_{\mu} a^{\mu})
\label{s7}
\eeq
or

\beq
Z_{Th}[b_{\mu}] = \exp(-\frac{N}{2g^2} \int \dv b_{\mu}b^{\mu})\times
Z_{SD}[b_{\mu}]
\label{s8}
\eeq
In order to connect this with the Maxwell-Chern-Simons system,
let us again consider Jackiw-Deser \cite{DJ} interpolating
action $S_I$ (eq.\ref{Master}) but now in the presence of sources:

\beq
S_I[a_{\mu},A_{\mu}; b_{\mu}] =
S_{I}[a_{\mu}, A_{\mu}] + \frac{\sqrt N}{g}\int \dv a_{\mu} b^{\mu}
\label{fin11}
\eeq
By integrating out $A_{\mu}$, one easily shows that the corresponding
partition function
$Z_{I}[b_{\mu}]$ coincides with  $Z_{SD}[b_{\mu}]$,

\beq
Z_{I}[b_{\mu}] = Z_{SD}[b_{\mu}]
\label{ag}
\eeq
Now, if one integrates first over  $a_{\mu}$ one has

\begin{eqnarray}
Z_I[b_{\mu}] &=&  \exp(\frac{N}{2g^2} \int \dv b_{\mu}b^{\mu})\times
\int DA_{\mu}
\exp( -\int \dv \frac{1}{4}F_{\mu \nu} F^{\mu \nu} + \nonumber \\
& &
 \pm i\frac{2\pi}{g^2}\int \dv
\epsilon_{\mu \nu \alpha}\partial^{\mu}A^{\nu} A^{\alpha}
+ \frac{\sqrt N}{g} \int \dv \epsilon_{\mu \nu \alpha}
\partial^{\mu}A^{\nu} b^{\alpha})
\label{fon2}
\end{eqnarray}
or
\beq
Z_I[b_{\mu}] =  \exp(\frac{N}{2g^2} \int \dv b_{\mu}b^{\mu}) \times
Z_{MCS}[b_{\mu}]
\label{inter}
\eeq
Finally, using eq.(\ref{s8}) we have the identity between partition functions
for the Thirring and Maxwell-Chern-Simons models in the presence of sources

\beq
Z_{Th}[b_{\mu}] \approx Z_{MCS}[b_{\mu}]
\label{fue}
\eeq
Here too, it is convenient to  rescale the vector potential $A_{\mu} \to
(g/\sqrt{4\pi}) A_{\mu}$, so that the MCS action takes the standard form.
With this change, $Z_{MCS}[b_{\mu}]$ in (\ref{fue}) now reads,

\begin{eqnarray}
Z_{MCS}[b_{\mu}]
=
\int DA_{\mu} \exp( &-&\int \dv \frac{1}{4e^2}F_{\mu \nu}
F^{\mu \nu} \pm i\frac{1}{2}\int \dv
\epsilon_{\mu\nu\alpha}\partial^{\mu}A^{\nu}A^{\alpha}
\nonumber \\
& +&\sqrt{\frac{N}{4\pi}}
\int \dv \epsilon_{\mu \nu \alpha}\partial^{\mu}A^{\nu} b^{\alpha})
\label{fon22}
\end{eqnarray}
where $e^2 = 4\pi/g^2$.

{}From eqs.(\ref{fue})-(\ref{fon22})  we see that the bosonization
rule for the fermion current reads, to leading order in $1/m$,

\beq
\bar\psi \gamma^{\mu} \psi  \to i{\sqrt \frac{N}{4\pi}}
\epsilon^{\mu \nu\alpha}
\partial_{\nu}A_{\alpha}
\label{aq}
\eeq
A few comments are in order. Firstly,
the bosonized expression for the fermion current
is manifestly conserved. Secondly, this formula
is the $2+1$-dimensional analog of the $1+1$-dimensional result
$\bar \psi \gamma^{\mu} \psi  \to {({1}/{\sqrt{\pi}})}
\epsilon_{\mu \nu} \partial^\nu \phi$. (The factor of $i$ in
the expression for the current in eq.(\ref{aq}) appears because,
here, we are working in Euclidean space). However, unlike
the $1+1$-dimensional formula, which is a short distance identity
valid for length scales long compared to a short distance cutoff
but small compared with the Compton wavelength of the fermion,
the $2+1$-dimensional identity is valid only for length scales
long compared with the Compton wavelength of the fermion.

We give now a first application of the bosonization formulas
and, in this way, explore their physical content. The effective
action of eq.(\ref{casi}) has a Chern-Simons term which
controls its long distance behavior. It is well
known\cite{Pol,witten} that the Chern-Simons gauge theory is a
theory of knot invariants which realizes the representations of
the Braid group. These knot invariants are given by expectation
values of Wilson loops in the Chern-Simons gauge theory. In
this way, it is found that the expectation values of the Wilson
loop operators imply the existence of excitations with
fractional statistics. Thus, it is natural to seek the
fermionic analogue of the Wilson loop operator $W_{\Gamma}$
which, in the Maxwell-Chern-Simons theory is given by
\beq
W_{\Gamma}=\langle \exp\{i {\frac{\sqrt N}{g}}\oint_{\Gamma}
A_\mu dx^\mu \}\rangle \label{wloop} \eeq
where $\Gamma$ is the
union of a an arbitrary set of closed curves (loops) in three
dimensional euclidean space. Given a closed loop (or union of
closed loops ) $\Gamma$, it is always possible to define a set
of open surfaces $\Sigma$ whose boundary is $\Gamma$, {\it
i.e.} $\Gamma= \partial \Sigma$. Stokes' theorem implies that

\begin{eqnarray} \langle \exp \{ i {\frac{\sqrt N}{g}}
\oint_{\Gamma} A_\mu dx^\mu \}\rangle &=& \langle \exp \{ i
{\frac{\sqrt N}{g}} \int_{\Sigma} dS_\mu \epsilon^{\mu \nu
\lambda} \partial_\nu A_\lambda \}\rangle \nonumber \\
&=&\langle \exp \{i {\frac{\sqrt N}{g}} \int \dv \epsilon^{\mu
\nu \lambda} \partial_\nu A_\lambda  \; b_\lambda \}\rangle
\label{stokes} \end{eqnarray}
is an identity. Here
$b_{\lambda}(x)$ is the vector field \beq b_\lambda(x)=
n_\lambda (x) \delta_{\Sigma}(x) \label{support} \eeq where
$n_\lambda$ is a field of unit vectors normal to the surface
$\Sigma$ and $\delta_\Sigma(x)$ is a delta function with
support on $\Sigma$. Using eq.(\ref{fue}) we find that this
expectation value becomes, in the Thirring Model, equivalent to

\beq W_{\Gamma}=\langle \exp \{ i {\frac{\sqrt
N}{g}}\int_{\partial \Sigma} dx_\mu A^\mu \}
\rangle_{MCS}=\langle \exp \{ \int_{\Sigma} dS_\mu {\bar \psi}
\gamma^\mu \psi \}\rangle_{Th} \label{thirringloop}
\eeq
More
generally we find that the Thirring operator ${\cal
W}_{\Sigma}$
\beq {\cal W}_{\Sigma}=\langle \exp \{ q
\int_{\Sigma} dS_\mu {\bar \psi} \gamma^\mu \psi \}\rangle_{Th}
\label{thirringop} \eeq
obeys the identity
\beq \langle \exp \{
q \int_{\Sigma} dS_\mu {\bar \psi} \gamma^\mu \psi
\}\rangle_{Th}= \langle \exp \{ i q{\frac{\sqrt N}{g}}
\oint_{\Gamma} A_\mu dx^\mu \}\rangle{MCS} \label{flux}
\eeq
for an arbitrary fermionic charge $q$.

The identity (\ref{flux}) relates the flux of the fermionic current
through an open surface $\Sigma$ with the Wilson loop operator
associated with the boundary $\Gamma$ of the surface.
The Wilson loop operator can be trivially calculated in the
Maxwell-Chern-Simons theory. For very large and smooth loops
the behavior of the Wilson loop operators is dominated by the
Chern-Simons term of the action. The result is a topological
invariant which depends only on the linking number $\nu_{\Gamma}$
of the set of curves $\Gamma$ \cite{Pol,witten}.
By an explict calculation one finds
\beq
\langle \exp \{ q \int_{\Sigma} dS_\mu {\bar \psi} \gamma^\mu \psi \}
\rangle_{Th}=
\exp \{ \mp i \nu_{\Gamma} {\frac{Nq^2}{8 \pi}}\}
\label{fracstat}
\eeq
This result implies that the non-local Thirring loop
operator ${\cal W}_{\Sigma}$ exhibits fractional statistics
with a statistical angle $\delta={{Nq^2}/{8 \pi}}$.
The topological significance of this result bears close resemblance
with the bosonization identity in $1+1$ dimensions between the
circulation of the fermion current on a closed curve and the
topological charge (or instanton number) enclosed in the interior
of the curve\cite{colemanlibro}. From the point of view of the
Thirring model, this is a most surprising result which reveals
the  power of the bosonization identities. To the best of our
knowledge, this is the first example of a purely fermionic
operator, albeit non-local, which is directly related to a
topological invariant.

In summary, in this work we presented a mappping between the low
energy sector of a self-interacting fermionic quantum field theory,
the massive Thirring model in $2+1$ dimensions, and a bosonic theory
of a  vector field, the Maxwell-Chern-Simons theory. This dual gauge
theory has a spectrum which consists of a spin one bosonic excitation
of mass $\pi/g^2$. We presented a number of identities for the
partition functions and for the generating function of the fermion current
correlation functions. The Wilson loop operator of the dual gauge
theory were found to have a natural expression in terms of the
fermion theory. As a byproduct, we found a fermion loop operator
which exhibits fractional statistics.

\vspace{1cm}
\underline{Acknowledgements}  This work was
supported in part by the National Science Foundation through the
grant NSF DMR-91-22385 at the University of Illinois at Urbana
Champaign (EF), by CICBA and CONICET (FAS) and by the NSF-CONICET
International Cooperation Program through the grant NSF-INT-8902032.
EF thanks the Universidad de La Plata for its kind hospitality.

\end{document}